\documentstyle[11pt,newpasp,twoside,epsf]{article}
\def\edcomment#1{\iffalse\marginpar{\raggedright\sl#1\/}\else\relax\fi}
\reversemarginpar

\begin{document}
\title{Unveiling Radio Loudness in Quasars}
\author{Cirasuolo M., Celotti A., Magliocchetti M., Danese L.}
\affil{SISSA/ISAS, via Beirut 4, 34014, Trieste, Italy}

\def\simlt{\mathrel{\rlap{\lower 3pt\hbox{$\sim$}}\raise 2.0pt\hbox{$<$}}}
\def\simgt{\mathrel{\rlap{\lower 3pt\hbox{$\sim$}} \raise
2.0pt\hbox{$>$}}}

\begin{abstract}
We present a new approach to tackle the issue of radio loudness in
quasars. 
We constrain a (simple) prescription for the intrinsic
distribution of radio-to-optical ratios by comparing properties of
Monte Carlo simulated samples with those of observed 
optically selected quasars. We find strong evidence for a dependence of the
radio luminosity on the optical one, even though with a large scatter.
The intrinsic distribution of the radio-to-optical ratios shows a peak at 
$R^*_{1.4} \sim 0.3$,  with only $\simlt 5$~ per cent of objects being included
in a high $R^*_{1.4}$ tail which identifies the radio loud regime.

\end{abstract}

\section{Introduction}
The origin of radio loudness of quasars is a long debated issue.
Radio observations of optically selected quasar samples showed only
10-40~\% of the objects to be powerful radio sources (Sramek \&
Weedman 1980; Condon et al. 1981; Miller, Peacock \&
Mead 1990; Kellermann et al. 1989). More interestingly, these early
studies suggested that quasars could be divided into the two
populations of ``Radio-Loud'' (RL) and ``Radio-Quiet'' (RQ) on the
basis of their radio emission.  Furthermore, Kellermann et al. (1989) found
that the radio-to-optical ratios, $R^*_{1.4}$ -- defined as the ratio
between radio (1.4 GHz) and optical (B band) rest frame luminosities --,
of these objects presented a bimodal distribution.  Miller, Peacock \&
Mead (1990) also found a dichotomy in the quasar population, although
this time radio luminosity was used as the parameter to define the
level of radio loudness.

In the last decade, our ability of collecting large samples of quasars
with faint radio fluxes has grown enormously, in particular thanks to
the FIRST (Faint Images of the Radio Sky at Twenty centimeters) Survey
at VLA (Becker, White \& Helfand 1995). However, despite the recent efforts,
radio loudness still remains an issue under debate. 
Works based on data from the FIRST survey (White et al. 2000;
Hewett et al. 2001) suggest that the found RL/RQ dichotomy could be
due to selection effects caused by the brighter radio and optical
limits of the previous studies.  On the contrary, Ivezic et al.
(2002) seem to find evidence for bimodality in a sample drawn from the Sloan
Digital Sky Survey (SDSS). More recently Cirasuolo et al.  (2003a)
-- analyzing a new sample obtained by matching
together the FIRST and 2dF QSO Redshift Survey -- ruled
out the classical RL/RQ dichotomy in which the distributions of
radio-to-optical ratios and/or radio luminosities show a deficit of
sources, suggesting instead a smoother transition between the RL and
the RQ regimes.

Clearly, the uncertainties on the presence of a dichotomy, the
character of radio loudness and the consequent poor knowledge of its
origin (dependence on BH mass, optical luminosity etc.) are due to the
analysis of different samples, often very inhomogeneous because of
selection effects both in the optical and radio bands, i.e. the lack
of a single sample covering all the ranges of optical and radio
properties of quasars.
\section{The model}
In order to shed some light on this issue, we adopted the
alternative approach of starting from simple assumptions on the
intrinsic properties of the quasar population as a whole -- namely an
optical quasar luminosity function and a prescription to associate a
radio power to each object - and, through Monte Carlo simulations,
generate unbiased quasar samples (Cirasuolo et al. 2003b). 
By applying observational limits in
redshift, apparent magnitude and radio flux we can then compare the
results of the simulations with the properties of observed samples.
The aim of this approach is of course twofold: constrain the initial
hypothesis on the intrinsic nature of quasars, by requiring
properties of the simulated samples -- such as $R^*_{1.4}$ and radio
power distributions, fraction of radio detections etc. -- to be in
agreement with the observed ones; test the effects of the
observational biases on each sample by simply changing the
observational limits. In order to cover a range as wide as possible 
of radio activity we
choose three samples of optically selected quasars for which radio data are
available, namely the 2dF Quasar Redshift
Survey (Cirasuolo et al. 2003a), the Large Bright Quasar Survey (Hewett et al.
2001) and the Palomar Bright Quasar Survey (Kellermann et al. 1989)

We decided to assume, as the two fundamental ingredients to describe the 
simulated quasar population, a well defined optical luminosity function 
obtained from the
2dF Quasar Redshift Survey (Croom et al. 2001)
- from which to obtain redshift and optical magnitude for the
sources - and different parameterizations for the distribution of 
radio-to-optical ratios which provide each source with a radio luminosity.
A solution, able to reproduce the properties of observed samples, has
been found by assuming radio and optical luminosities to be related to each
other even
though with a large scatter. The radio-to-optical ratio and radio
power distributions -- modeled as two gaussians and 
corresponding to this solution -- 
are displayed in Figure and the model parameters are given in Table.

\begin{table}
\begin{center}
\begin{tabular}{ccccc} \hline \hline
 $x_1$          & $\sigma_1$    & $x_2$ & $\sigma_2$ & Fraction\\ \hline
 $ 2.7 \pm 0.2$ & $0.7 \pm 0.2$ & $-0.5 \pm 0.3$ & $0.75 \pm 0.3$ &$97 \pm 2$~per cent\\
\hline \hline
\end{tabular}
\end{center}
\caption{\label{tab:mc} Best-fit parameters for the  model, expressed in
$\log_{10}\: R^*_{1.4}$. $x_1$ and $\sigma_1$ are the center and
dispersion of the Gaussian in the RL regime, while $x_2$ and
$\sigma_2$ are those for the Gaussian in the RQ one. ``Fraction''
indicates the percentage of objects having radio-to-optical ratios described by 
the second Gaussian.}
\end{table}
\begin{figure}
\center{{
\plotfiddle{shape_R_P.ps}{7cm}{0}{40}{40}{-100}{-60}
}}
\caption{\label{shape_R_P} Distribution of radio-to-optical ratios
(top panel) and radio powers (bottom panel) obtained from the best-fit 
set of parameters (see Table). 
The distributions are plotted in a
binned form and the shaded regions indicate the range of $R^*_{1.4}$
and $P_{1.4}$ for which no data are available.}
\end{figure}

\section{Discussion}
The first point worth stressing is the ``uniqueness'' of the
solution found. The combination of all the observational constraints
is very cogent and thus, despite large errors on each constraint, we
find that only one set of parameters is able to simultaneously reproduce all
measurements from the three surveys. Furthermore, 
the uncertainties associated to the various parameters are in this case 
relatively small (see Table).  \\ 
It is important to remark here that
in order to reproduce the
data we need a dependence of the radio luminosity on the optical one,
even though with a large scatter. In particular, the successful model
accounts for the dependence of the observed fraction of radio
detected quasars on apparent and absolute optical magnitudes, as due to
selection effects.\\
Given the uniqueness of the solution, the main result of this work is
indeed the fact that we can put rather tight constraints on the
intrinsic radio properties of quasars.  The distributions shown in the 
Figure  could then describe the unbiased view of the
properties of the whole quasar population and this might possibly help
us to understand the physical mechanism(s) responsible for radio
emission.  First of all, in the $R^*_{1.4}$ distribution we note no
lack or deficit of sources between the RL and RQ regimes: the
distribution has a peak at $R^*_{1.4} \sim 0.3$ and decreases
monotonically beyond that value with only 
a small fraction ($\simlt 5$~per cent) of objects
found in the RL regime which represent the long flat tail of the total
distribution. This result contrasts the classical view of a
RL/RQ dichotomy where a gap separates the two populations.
Nevertheless we can still talk about a ``dichotomy'' in the sense that
the data are compatible with an asymmetric distribution, with a steep
transition region and  only a small fraction of sources having
high values of $R^*_{1.4}$. This result is in agreement, within the errors, 
with the findings from the new analysis of the 
SDSS presented by Ivezic et al. during this 
conference. They still claim the presence of a local minimum dividing the
two populations, even though this is now less pronounced than what claimed by  
Ivezic et al. (2002).
While the advantage of SDSS is clearly the large statistics, it is also
limited to the RL regime by the 1 mJy cut of the FIRST
Survey.
On the other hand our method allows to explore a wider range in radio loudness. 
In any case our findings are consistent, within the errors on best-fit 
parameters, with the presence of  such a shallow minimum where the two
gaussians -- which describe the $R^*_{1.4}$ distribution -- cross each other.
The RL regime would simply remain a long flat tail of the asymmetric 
$R^*_{1.4}$ distribution and it would be clearly of great interest to apply 
our analysis directly to the SDSS data once its completeness is achieved. 

\references
Becker R.H., White R.L., \& Helfand D.J., 1995, \apj, 450, 559 \\
Cirasuolo M., Magliocchetti M., Celotti A., \& Danese L., 2003a, \mnras, 341,
 993 \\
Cirasuolo M., Celotti A., Magliocchetti M., \& Danese L., 2003b, \mnras, 346,
447 \\
Condon J.J., O'Dell S.L., Puschell J.J., \& Stein W.A. 1981, \apj, 246, 624 \\
Croom S.M., Smith R.J., Boyle B.J., Shanks T., Loaring N.S., Miller L.,
\&  Lewis I.J., 2001, \mnras, 322, L29 \\
Hewett P.C., Foltz C.B., Chaffee F.H., 2001, \aj, 122, 518 \\
Ivezic Z., et al., 2002, \aj,
124, 2364 \\
Kellermann K.I., Sramek R., Schmidt M., Shaffer D.B., \& Green R, 1989, \aj,
 98, 1195 \\
Miller L., Peacock J.A., \& Mead A.R.G., 1990, \mnras, 244, 207 \\
Sramek R.A., \& Weedman D.W., 1980, \apj, 238, 435 \\
White R.L., et al., 2000,  \apjs, 126, 133
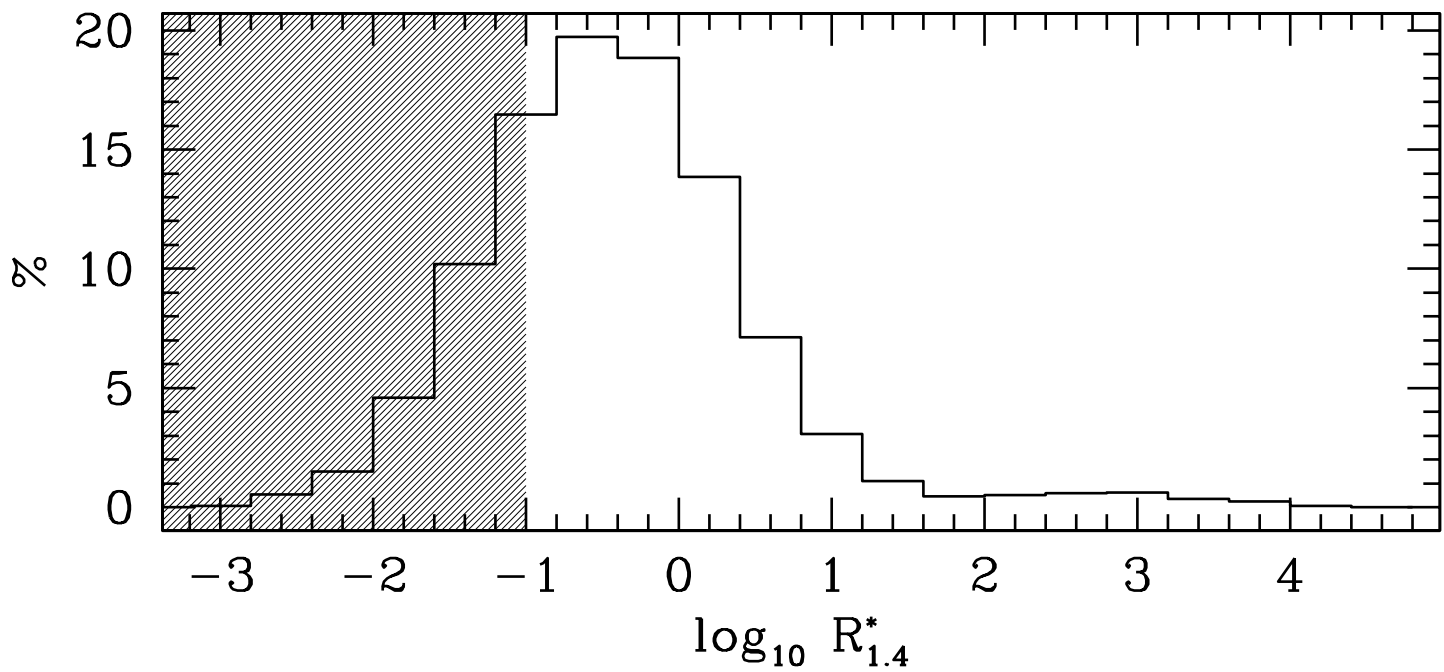
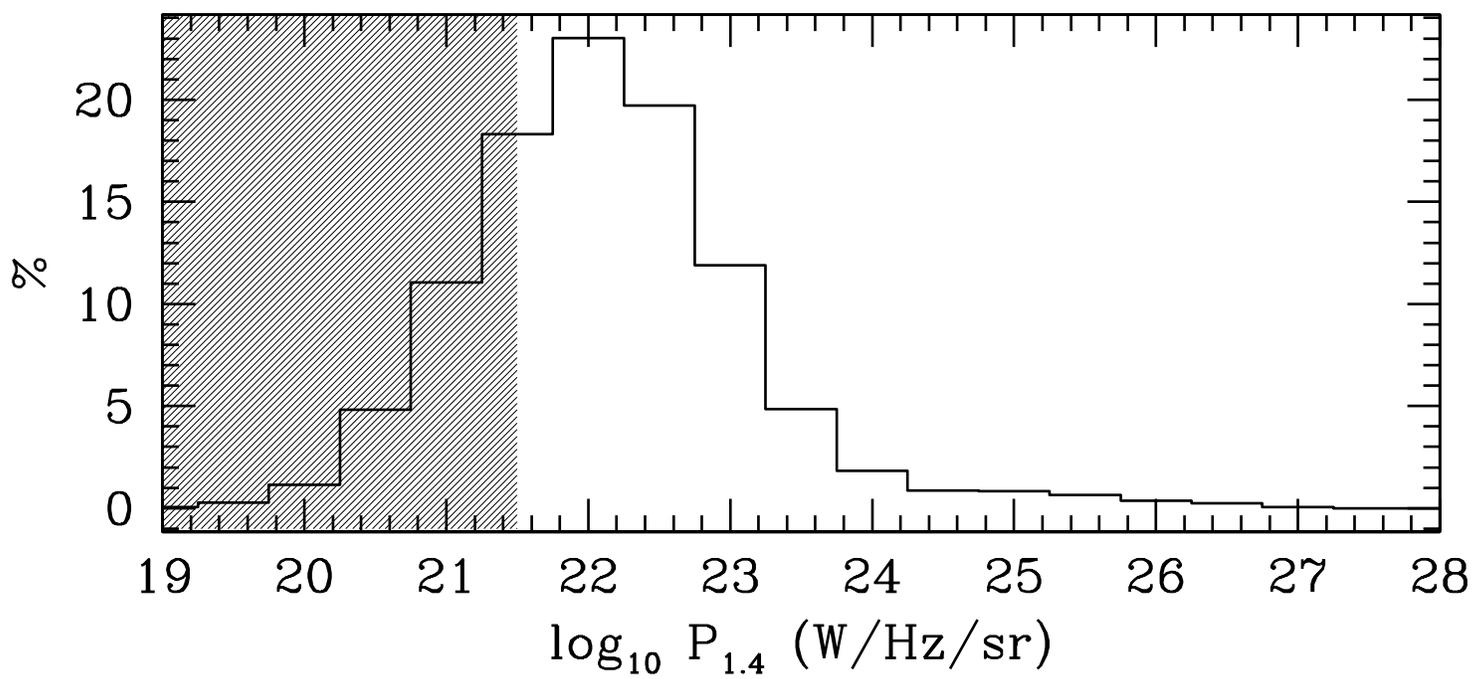
\end{document}